\documentclass[a4paper,twoside]{article}
\usepackage{amssymb}
\usepackage{amsmath}
\usepackage{amsfonts}

\begin{document}

\title{\Large\bf{Quantum Gravitational Collapse
}} 

\author{\\ Leonardo
 Modesto\\ 
 \em\small{Centre de Physique Th\'eorique de Luminy,}\\
\em\small{Universit\'e de la M\'editerran\'ee, F-13288 Marseille, EU}\\[-1mm]
 }

\date{\ } 
\maketitle

\begin{abstract}
We apply the recent results in Loop Quantum Cosmology and in the resolution of Black Hole
singularity to the gravitational collapse of a star. We study the dynamic of the space time in the interior of the Schwarzschild radius. 
In particular in our simple model we obtain the evolution of the matter inside 
the star and of the gravity outside the region where the matter is present. The boundary 
condition identify an unique time inside and outside the region where the matter is present.
We consider a star during the collapse in the particular case in which   
inside the collapsing star we take null pressure, homogeneity and isotropy. The space-time
outside the matter is homogeneous and anisotropic. We show that the space time is 
singularity free and that we can extend dynamically the space-time beyond the classical singularity.
\end{abstract}

\section*{Introduction}

We consider the gravitational collapse in the particular case 
the spatial section of space time is subdivided in two parts, a first part 
with matter and a second part without. 
Classically we consider a dust star of null pressure and we report 
the known results for the gravitational collapse to produce a
classical singularity. In the quantum theory we use the recent 
results obtained for the Schwarzschild solution inside the horizon \cite{M2}
to describe the quantum theory outside the region which contains the matter.
In fact outside the matter and inside the horizon we consider a 
general two dimensional minisuperspace with space sections of topology 
$\mathbf{R} \times \mathbf{S}^2$ which is know as Kantowski-Sachs space
time \cite{KS}. The Schwarzschild space time inside the horizon is a particular 
representative of this class of metrics. 
On the other side inside the matter we take the Friedmann space time with 
positive curvature. 
In this paper we use the same non Schr$\ddot{\mbox{o}}$dinger procedure 
of quantization used in the previous paper \cite{work1} and in the work 
of V. Husain and O. Winkler on quantum cosmology and introduced by 
Halvorson \cite{Fonte.Math} and also by A. Ashtekar, S. Fairhust and
J. Willis \cite{AFW}. 
In particular in this paper we resolve the Hamiltonian constraint inside 
and outside the region where the matter is located. We resolve the
Hamiltonian constraint inside the matter without specifing the matter type 
but in the case of dust matter we match the wave function inside and 
outside the matter obtaining a temporal coordinate for all the space 
time that is connected with the star radius. 
The wave function "$\psi(\tilde{a}, \phi, a)$" that resolves 
the Hamiltonian constrain is interpreted as the wave function for the 
matter field "$\phi$" (which represents the dust matter) 
and the gravitational field "$\tilde{a}$" at the time "$a$";
"$a$"  is proportional to the radius of the star. 
Than we consider the particular case of a scalar field and 
we quantize the complete Hamiltonian constraint  using the same 
quantization procedure \cite{work1}, \cite{AFW}, \cite{KS}
also for the matter field. In this case we cannot match the
Schwarzschild space time outside the matter because for the
scalar field the pressure is not equal to zero. On the other side
in this case we can match with the Vaydia space time as in the
paper \cite{BojBH}. 

The main result is that the space time is singularity free. In fact 
using the method \cite{Thie} we can define the inverse volume density 
and the operator $\frac{1}{a^6}$ in terms of the holonomy analog and 
the volume itself and we show that these quantities are always finite and 
upper bounded. The operator  $\frac{1}{a^6}$ is proportional to 
$\mathcal{R}_{\mu \nu \rho \sigma} \, \mathcal{R}^{\mu \nu \rho \sigma}$ 
in the case of the Schwarzschild metric and is divergent in the classical theory
producing the singularity. From the point of view of the matter region 
we show that the classical singular quantity "$1/a$" is finite in the quantum theory.
Another important result is that using also the result in \cite{Thie} we can
obtain the Hamiltonian constraint in terms of the volume and at the quantum 
level we have a discrete equation depending on three parameters: 
the matter field "$\phi$" (dust matter) , the gravitational field "$\tilde{a}$" 
and the time "$a$". This equation says that we can extend dynamically 
the space time beyond the classical singularity.

The paper is organized as follows : in the first section we resume the classical collapse
inside and outside the horizon in the case of dust matter. 
In the second section we introduce the Hamiltonian constraint for gravity inside the horizon
in the two region inside and outside the matter.
In the third section we quantize the system using the non Schr$\ddot{\mbox{o}}$dinger procedure of quantization \cite{Fonte.Math} and \cite{AFW}.
In this section we show that the inverse volume operator is singularity free in quantum gravity 
and also that the Hamiltonian constraint acts like a difference operator as in loop quantum cosmology.
In the fourth section we quantize an homogeneous and isotropic scalar field using also in
this case the analog of  loop quantization. In this case we cannot match with the
the Kantowski-Sachs Space-Time because the pressure is not equal to zero.

\section{Classical Theory}

\subsection{Classical Gravitational Collapse}

The space-time outside the matter but inside the horizon is an homogeneous, anisotropic 
space-time with spatial section of topology $\mathbf{R} \times \mathbf{S}^2$, this is the Kantowski-Sachs Space-Time that we have studied in \cite{M2}. The metric in this case is 
\begin{eqnarray}
ds^2 = - \tilde{N}(t) dt^2 + \tilde{a}^2 (t) dr^2 + b^2 (t)  (\sin^2 \theta d\phi^2 + d \theta^2).
\label{metric.out}
\end{eqnarray}
On the other side we assume that inside the matter the space-time is homogeneous, isotropic
and with null pressure,  so the metric must to be the Friedmann solution
\begin{eqnarray}
ds^2 = - N(t) dt^2 + a^2 (t) \big[d\chi^2 + \sin^2 \chi (\sin^2 \theta d\phi^2 + d \theta^2)\big].
\label{metric.in}
\end{eqnarray}
Usually in the classical theory the much simple example is the gravitational collapse of a 
spherically symmetric, perfect fluid star of zero pressure and uniform density inside the region 
where the matter is localized and the Schwarzschild solution outside the matter. The metric
(\ref{metric.out}) reduces to the Schwarzschild solution for the particular case 
$\tilde{a}^2 = \frac{2 M G_n}{b} -1$ where $M$ is the ADM mass of the black hole. 
In this particular case we can match the two space time regions and in particular
we can match the two dimensional spherical surface inside and the two dimensional
surface outside the matter.
We will extend the match also at the quantum level.

\subsection{The Hamiltonian Constraint}

The $Diff$-constraints for the metrics in (\ref{metric.out}) and (\ref{metric.in}) are identically satisfied.
The Hamiltonian constraint outside the region where is localized the matter is calculated in \cite{M2} and it is 
\begin{eqnarray}
H_{out} = \frac{ G_N \, |{\tilde{a}}| \, p_{\tilde{a}}^2}{2R \, b^2} - \frac{G_N p_{\tilde{a}} p_b \, \mbox{sgn}(\tilde{a})}{ R b} - \frac{R}{2 G_N} |{\tilde{a}}| ;
\label{HC.Mini} 
\end{eqnarray}
the Hamiltonian constraint inside the matter is 
\begin{eqnarray}
H_{in} = -\Big(\frac{p_a^2}{8 |a|} + 2 |a| \Big) + \frac{16 \pi G_N}{3} H_\phi(a),
\label{HC.in} 
\end{eqnarray}
where $H_{\phi}(a)$ is the hamiltonian constraint of the dust ({\em first case}), in this case
we can match with the Schwarzschild solution or of an isotropic and homogeneous
scalar matter ({\em second case}) but in this case we cannot match with the Schwarzschild solution. The volume of the space section inside and outside the matter are respectively 
\begin{eqnarray}
&& V_{in} = \int_0^{\chi_0} d\chi \, \int_0^{2 \pi} d \phi \, \int_{0}^{\pi} d \theta \, h^{1/2}_{in} = 
 2 \pi\big(\chi_0 - \sin(\chi_0) \, \cos(\chi_0)\big) \, |a|^3 \equiv \mathcal{V}(\chi_0) \, |a|^3 \nonumber \\
&& V_{out} = \int_0^{R} dr \, \int_0^{2 \pi} d \phi \, \int_{0}^{\pi} d \theta \, h^{1/2}_{out} = 4 \pi R |\tilde{a}| b^2 
\label{Volume}
\end{eqnarray}
where $\chi_0$ define the boundary of the star of radius $R_0 = |a| \chi_0$ and where $R$ is a cut-off on the space radial coordinate. We can work also with radial densities 
because the model is homogeneous and all the following results remain identical. In another way, 
the spatial homogeneity enables us to fix a linear radial cell $\mathcal{L}_r$ and restricts all 
integrations to this cell \cite{Boj}.   
We have two canonical pairs outside the matter, one is given by $\tilde{a} \equiv x_{\tilde{a}}$ and $p_{\tilde{a}}$, the other is given by $b \equiv x_b$ and $p_b$ for which the Poisson brackets are $\{x_{\tilde{a}},p_{\tilde{a}}\} = 1$ and $\{x_b,p_b\} = 1$. Inside the matter we have two canonical couple $x_a$ and $p_a$ with $\{x_a,p_a\} = 8 \pi G_N$ and the canonical couple for the matter
that for the scalar field consists of $\phi$ and $p_{\phi}$.
From now on we consider $x_a, x_b, x_{\phi}, x_{\tilde{a}} \in \mathbb{R}$ and we will introduce the modulus of $x_a$, $x_b$, $x_{\phi}$ and $x_{\tilde{a}}$ where it is necessary.

\subsection{Classical Theory Inside the Matter}
We study the gravity sector and the matter sector. In the two cases we  
follow \cite{work1}, \cite{Fonte.Math}, \cite{Boj} and \cite{AFW}. We introduce an algebra of classical observable and we write the quantities of physical interest in terms of these variables. \\
\leftline{\underline{\sl Gravity Sector}} 
We start with the gravity sector and as in Loop Quantum Gravity we use the fundamental 
variables $x_a, U_{\gamma_a}$ and the volume $V_{in} = \mathcal{V}(\chi_0) \, |x_a|^3$, where  
\begin{eqnarray}
&& U_{\gamma_a}(p_a) \equiv \mbox{exp} \Big(\frac{i \gamma_a}{L_a} \,  p_a\Big), \nonumber \\
\label{Ugammaa}
\end{eqnarray}   
the parameter $\gamma_a$ is a real and $L_a$ fixes the unit of length \cite{AFW}. This variable can be seen as the momentum analog of the holonomy variable of loop quantum gravity.\\
We have also that 
\begin{eqnarray}
&& \{x_a, U_{\gamma_a}\} = i \frac{8 \pi G_N \gamma_a}{L_a} \, U_{\gamma_a}, \nonumber \\
&& U_{\gamma_a}^{-1} \{ V^n_{in} , U_{\gamma_a} \} = i \, \frac{24 \pi G_N \gamma_a}{L_a} \, n \, 
|x_a|^{3n -1} \mbox{sgn}(x_a) \mathcal{V}^n(\chi_0). \nonumber \\
\label{Poisson.Volume}
\end{eqnarray}
From this relations we can construct the following quantity that we will use extensively 
\begin{eqnarray}
&& \frac{\mbox{sgn}(x_a)}{\sqrt{|x_a|}} = - \frac{2 L_a i}{8 \pi G_N \gamma_a 
\mathcal{V}^{1/6}(\chi_0)} \, U_{\gamma_a}^{-1} \{ V^{\frac{1}{6}}_{in} , U_{\gamma_a} \}.
\label{bsua}
\end{eqnarray}
We are interested to the quantity $\frac{1}{|x_a|}$ because classically this 
quantity diverge because it is proportional to the Ricci invariant. 

We are also interested to the Hamiltonian constraint and to the dynamics of the minisuperspace model. \\
\leftline{\underline{\sl Matter Sector (first case)}} 
The first possibility 
is to take 
the dust matter with zero pressure, in this case the super momentum is zero
and the Hamiltonian constraint \cite{Brown} is very simple (see the appendix)
\begin{eqnarray}
H_{\phi} = |p_{\phi}|.
\label{Dust}
\end{eqnarray}
The canonical pair is $(\phi, p_{\phi})$ with Posson bracket is $\{\phi, p_{\phi}\} = 1$; as in 
the gravity sector and in analogy with loop quantum gravity we take as fundamental
variables $\phi$ and $U_{\gamma_{\phi}}$, where 
\begin{eqnarray}
&& U_{\gamma_{\phi}}(p_{\phi}) \equiv \mbox{exp} 
\Big(\frac{i \gamma_{\phi} \, 8 \pi G_N}{L_{\phi}} \,  p_{\phi}\Big), \nonumber \\
&& \{x_{\phi}, U_{\gamma_{\phi}}\} = i \frac{\gamma_{\phi} \, 8 \pi G_N}{L_{\phi}^2} \, U_{\gamma_{\phi}}
\label{UgammaphiDust}
\end{eqnarray}   
\leftline{\underline{\sl Matter Sector (second case)}} 
Another possibility is to take a scalar, isotropic and homogeneous scalar field with 
$H_{\phi}$ of the constraint equation (\ref{HC.in}) equal to 
\begin{eqnarray}
H_{\phi} = \frac{1}{2 |a|^3} \Big[p_{\phi}^2 + a^6 \, U(\phi)\Big].
\label{Scal.Con}
\end{eqnarray}
Also in this case the canonical pair is $(\phi, p_{\phi})$ with Posson bracket $\{\phi, p_{\phi}\} = 1$;  as fundamental variables we take $\phi$ and $U_{\gamma_{\phi}}$, where 
 \begin{eqnarray}
&& U_{\gamma_{\phi}}(p_{\phi}) \equiv \mbox{exp} \Big(\frac{i \gamma_{\phi} \, \sqrt{8 \pi G_N}}{L_{\phi}^2} \,  p_{\phi}\Big), \nonumber \\
&& \{x_{\phi}, U_{\gamma_{\phi}}\} = i \frac{\gamma_{\phi} \, \sqrt{8 \pi G_N}}{L_{\phi}^2} \, U_{\gamma_{\phi}}
\label{Ugammaphi}
\end{eqnarray}   

In the two cases using the introduced quantities we can quantize the matter sector as the gravity sector following \cite{AFW}.

\subsection{Classical Theory Outside the Matter}
The classical theory outside the matter but always inside the horizon has been described 
in reference \cite{M2}; we recall very rapidly the Hamiltonian constraint and the volume. 
The Hamiltonian constraint is 
 \begin{eqnarray}
&& H_c =\frac{ G_N \, |\tilde{a}| \, p_{\tilde{a}}^2}{2R \, b^2} - \frac{G_N p_a p_b \, \mbox{sgn}(\tilde{a})}{ R \, b} - \frac{R}{2 G_N} |\tilde{a}|, \nonumber \\
&&  V = 4 \pi R |x_{\tilde{a}}| \, |x_b|^2
.\label{HC.Mini2} 
\end{eqnarray}
where $R$ has been defined previously. The fundamental variables are $x_{\tilde{a}}, x_b$ and 
\begin{eqnarray}
&& U_{\gamma_{\tilde{a}}}(p_{\tilde{a}}) \equiv \mbox{exp} \Big(\frac{8 \pi G_N \gamma_{\tilde{a}}}{L_{\tilde{a}}^2} \, i \,  p_{\tilde{a}}\Big),  \nonumber \\
&& U_{\gamma_b}(p_b) \equiv \mbox{exp} \Big(\frac{8 \pi G_N \gamma_b}{L_b} \, i \,  p_b\Big), 
\label{UaUb}
\end{eqnarray}   
As in the previews paper \cite{M2}, we will use the following relations to quantize the
sistem: 
\begin{eqnarray}
&& \{x_{\tilde{a}} , U_{\gamma_{\tilde{a}}}(p_{\tilde{a}})\} = 8 \pi G_N \frac{i \, \gamma_{\tilde{a}}}{L_{\tilde{a}}^2} U_{\gamma_{\tilde{a}}}(p_{\tilde{a}}), \nonumber \\
&& \{x_b , U_{\gamma_b}(p_b)\} = 8 \pi G_N \frac{i \, \gamma_b}{L_b} U_{\gamma_b}(p_b), \nonumber \\
&& U_{\gamma_{\tilde{a}}}^{-1} \{ V^m , U_{\gamma_{\tilde{a}}} \} = (4 \pi R |x_b|^2)^m \, m \, |x_{\tilde{a}}|^{m-1} \mbox{i} \gamma_{\tilde{a}} \frac{8 \pi G_N}{L_{\tilde{a}}^2} 
\mbox{sgn}(x_{\tilde{a}}), \nonumber \\
&& U_{\gamma_b}^{-1} \{ V^n , U_{\gamma_b} \} = (4 \pi R |x_{\tilde{a}}|)^n \, 2 n \, |x_b|^{2n-1} \mbox{i} \gamma_b \frac{8 \pi G_N}{L_b} \mbox{sgn}(x_b).
\label{Poisson.Volume1}
\end{eqnarray}
With the quantities (\ref{Poisson.Volume1}) we construct the Hamiltonian constraint and 
the classically singular observable.

\section{Quantum Theory}

We construct the quantum theory in analogy with the procedure used in loop quantum gravity and in particular following \cite{work1}. First of all for any canonical couple inside and outside the matter we must take an algebra of classical functions that is represented as quantum configuration operators. For any couple of canonical variable we choose the algebra generated by the function
\begin{eqnarray}
W(\lambda) = e^{i \lambda x/L}
\end{eqnarray}
where $\lambda \in \mathbb{R}$. The algebra consists of all function of the form 
\begin{eqnarray}
f(x) = \sum_{j=1}^{n} c_j e^{i \lambda_j x/L}
\end{eqnarray}
where $c_j \in \mathbb{C} $ and their limits with respect to the supremum norm and with $x$ we
indicate $x_a$ or $x_b$. This algebra is the $algebra$ $of$ $almost$ $periodic$ $function$ $over$ $\mathbb{R}$ ($AP(\mathbb{R})$). The algebra ($AP(\mathbb{R})$) is isomorphic to $C(\bar{\mathbb{R}}_{Bohr})$ that is the algebra of continuous functions on the Bohr-compactification of $\mathbb{R}$. 
This suggests to take inside and outside the matter region the Hilbert space $L_2(\bar{\mathbb{R}}_{Bohr}^2, d \mu_0)$, where $d \mu_0$ is the Haar measure on $\bar{\mathbb{R}}_{Bohr}^2$.\\
\leftline{\underline{\sl Inside the Matter (first case)}} 
We recall that inside the matter the canonical couples are the matter canonical couple
$(\phi, p_{\phi})$ and the gravity canonical couple $(a, p_a)$. 
The complete Hilbert space inside the matter is the tensor product  
\begin{eqnarray}
&& |\lambda_a \rangle \otimes  |\lambda_{\phi} \rangle \equiv | e^{i \lambda_a \, x_a / L_a}\rangle \otimes
       |e^{i \lambda_{\phi} x_{\phi}/L_{\phi}} \rangle, \nonumber \\
&& \langle \mu_a | \lambda_a \rangle = \delta_{\mu_a, \lambda_a} \hspace{1cm} 
      \langle \mu_{\phi} | \lambda_{\phi} \rangle = \delta_{\mu_{\phi}, \lambda_{\phi}}. 
      \label{base.in.dust}
      \end{eqnarray}
The action of the configuration operators $\widehat{W}_a(\lambda_a)$ and $\widehat{W}_{\phi}(\lambda_{\phi})$ on the base is definited by 
\begin{eqnarray}
&& \widehat{W}_a(\lambda_a) | \mu_a \rangle = e^{i \lambda_a \hat{x}_a/L_a} | \mu_a \rangle = e^{i \lambda_a \mu_a } |\mu_a \rangle, \nonumber \\
&& \widehat{W}_{\phi}(\lambda_{\phi}) | \mu_{\phi} \rangle = e^{i \lambda_{\phi} 
 \hat{x}_{\phi}/L_{\phi}} | \mu_{\phi} \rangle = e^{i \lambda_{\phi} \mu_{\phi} } 
|\mu_{\phi} \rangle.
\end{eqnarray}
Those operators are weakly continuous in $\lambda_a$,  $\lambda_b$ and this imply the existence of  self-adjoint operators $\hat{x}_a$ and $\hat{x}_b$, acting on the basis states according to  
\begin{eqnarray}
&& \hat{x}_a |\mu_a \rangle = \mu_a L_{a}|\mu_a \rangle,  \nonumber \\
&& \hat{x}_{\phi} |\mu_{\phi} \rangle =  \mu_{\phi} L_{\phi} |\mu_{\phi} \rangle.
\label{xoperator1P}
\end{eqnarray}
Now we introduce the operators corresponding to the classical momentum functions 
$U_{\gamma_a}$ and $U_{\gamma_{\phi}}$ of (\ref{Ugammaa}) and (\ref{Ugammaphi}). 
We define the action of $\hat{U}_{\gamma_a}$ and $\hat{U}_{\gamma_{\phi}}$ on the basis states using the definitions (\ref{xoperator1}) and using a quantum analog of the Poisson brackets between $x_a$ and $U_{\gamma_a}$ and $x_{\phi}$ and $U_{\gamma_{\phi}}$ 
\begin{eqnarray}
&& \hat{U}_{\gamma_a} |\mu_a \rangle = | \mu_a - \gamma_a \rangle \hspace{1cm} 
      \hat{U}_{\gamma_{\phi}} |\mu_{\phi} \rangle = | \mu_{\phi} - \gamma_{\phi} \rangle,    
 \nonumber \\ 
&& \left[ \hat{x}_a , \hat{U}_{\gamma_a} \right] = - \gamma_a \, L_a \, \hat{U}_{\gamma_a} \hspace{1.5cm}
       \left[ \hat{x}_{\phi} , \hat{U}_{\gamma_{\phi}} \right] = 
       - \gamma_{\phi} L_{\phi} \hat{U}_{\gamma_{\phi}}.
       \end{eqnarray} 
\leftline{\underline{\sl Inside the Matter (second case)}} 
In this second case we consider the scalar matter inside the horizon. The canonical couples are the matter canonical couple $(\phi, p_{\phi})$ and the same gravity canonical couple $(a, p_a)$ as in the
case of dust matter. The Hilbert space is 
\begin{eqnarray}
&& |\lambda_a \rangle \otimes  |\lambda_{\phi} \rangle \equiv | e^{i \lambda_a \, x_a / L_a}\rangle \otimes
       |e^{i \lambda_{\phi} \sqrt{8 \pi G_N} x_{\phi}} \rangle, \nonumber \\
&& \langle \mu_a | \lambda_a \rangle = \delta_{\mu_a, \lambda_a} \hspace{1cm} 
      \langle \mu_{\phi} | \lambda_{\phi} \rangle = \delta_{\mu_{\phi}, \lambda_{\phi}}. 
      \label{base.in}
      \end{eqnarray}
The action of the configuration operators $\widehat{W}_a(\lambda_a)$ and $\widehat{W}_{\phi}(\lambda_{\phi})$ on the base is definited by 
\begin{eqnarray}
&& \widehat{W}_a(\lambda_a) | \mu_a \rangle = e^{i \lambda_a \hat{x}_a/L_a} | \mu_a \rangle = e^{i \lambda_a \mu_a } |\mu_a \rangle, \nonumber \\
&& \widehat{W}_{\phi}(\lambda_{\phi}) | \mu_{\phi} \rangle = e^{i \lambda_{\phi} 
\sqrt{8 \pi G_N} \hat{x}_{\phi}} | \mu_{\phi} \rangle = e^{i \lambda_{\phi} \mu_{\phi} } 
|\mu_{\phi} \rangle.
\end{eqnarray}
Those operators are weakly continuous in $\lambda_a$,  $\lambda_b$ and this imply the existence of  self-adjoint operators $\hat{x}_a$ and $\hat{x}_b$, acting on the basis states according to  
\begin{eqnarray}
&& \hat{x}_a |\mu_a \rangle = \mu_a L_a|\mu_a \rangle,  \nonumber \\
&& \hat{x}_{\phi} |\mu_{\phi} \rangle =  \frac{\mu_{\phi}}{\sqrt{8 \pi G_N}} |\mu_{\phi} \rangle.
\label{xoperator1}
\end{eqnarray}
The operators corresponding to the classical momentum functions 
are $\hat{U}_{\gamma_a}$ and $\hat{U}_{\gamma_{\phi}}$ and the action on the basis states is 
\begin{eqnarray}
&& \hat{U}_{\gamma_a} |\mu_a \rangle = | \mu_a - \gamma_a \rangle \hspace{1cm} 
      \hat{U}_{\gamma_{\phi}} |\mu_{\phi} \rangle = | \mu_{\phi} - \gamma_{\phi} \rangle,    
 \nonumber \\ 
&& \left[ \hat{x}_a , \hat{U}_{\gamma_a} \right] = - \gamma_a \, L_a \, \hat{U}_{\gamma_a} \hspace{1.5cm}
       \left[ \hat{x}_{\phi} , \hat{U}_{\gamma_{\phi}} \right] = 
       - \frac{\gamma_{\phi}}{\sqrt{8 \pi G_N}} \hat{U}_{\gamma_{\phi}}.
       \end{eqnarray} 
\leftline{\underline{\sl Outside the Matter}} 
Outside the matter the Hilbert space \cite{M2} is generated by the basis 
\begin{eqnarray}
&& |\lambda_{\tilde{a}} \rangle \otimes  |\lambda_b \rangle \equiv |e^{i \lambda_{\tilde{a}} x_{\tilde{a}}} \rangle \otimes
       |e^{i \lambda_b x_b/L_b} \rangle, \nonumber \\
&& \langle \mu_{\tilde{a}} | \lambda_{\tilde{a}} \rangle = \delta_{\mu_{\tilde{a}}, \lambda_{\tilde{a}}} \hspace{1cm} 
      \langle \mu_b | \lambda_b \rangle = \delta_{\mu_b, \lambda_b}. 
      \label{base.out}
      \end{eqnarray}
The self-adjoint operators $\hat{x}_{\tilde{a}}$ and $\hat{x}_b$ act on the basis states 
according to  
\begin{eqnarray}
&& \hat{x}_{\tilde{a}} |\mu_{\tilde{a}} \rangle = \mu_{\tilde{a}} |\mu_{\tilde{a}} \rangle,  \nonumber \\
&& \hat{x}_b |\mu_b \rangle = L_b \mu_b |\mu_b \rangle.
\label{xoperator}
\end{eqnarray}
The analog of the classical momentum functions 
$U_{\gamma_a}$ and $U_{\gamma_b}$ of (\ref{UaUb}) are 
 $\hat{U}_{\gamma_a}$ and $\hat{U}_{\gamma_b}$. Those operators act on the basis states 
\begin{eqnarray}
&& \hat{U}_{\gamma_{\tilde{a}}} |\mu_{\tilde{a}} \rangle = | \mu_{\tilde{a}} - \gamma_{\tilde{a}} \rangle \hspace{1cm} 
      \hat{U}_{\gamma_b} |\mu_b \rangle = | \mu_b - \gamma_b \rangle,    
 \nonumber \\ 
&& \left[ \hat{x}_{\tilde{a}} , \hat{U}_{\gamma_{\tilde{a}}} \right] = - \gamma_{\tilde{a}} \hat{U}_{\gamma_{\tilde{a}}} \hspace{1.5cm}
       \left[ \hat{x}_b , \hat{U}_{\gamma_b} \right] = - \gamma_b \hat{U}_{\gamma_b}.
       \end{eqnarray} 
Using the standard quantization procedure $[ \, , \, ] \rightarrow i \hbar \{ \, , \, \}$, and using the
first of the two equations of (\ref{Poisson.Volume}) and the second of (\ref{Ugammaphi}) we obtain 
\begin{eqnarray}
&& L \equiv L_{\tilde{a}} = L_b = L_a = L_{\phi} = \sqrt{8 \pi G_n \hbar}.
\label{Uphi}
\end{eqnarray}

\subsection{Singularity Resolution in Quantum Theory}
We resume in this section the singularity resolution of the gravitational collapse in loop
quantum gravity. In particular we report the regular spectrum of the operator $1/\mbox{det(E)}$
outside the matter which is connected with the Schwarzschild singularity and the 
operator $1/a$ which is connected with the singularity inside the matter. We can observe 
also that the singularity inside the matter is analogues to the cosmological singularity. 

We start with the space-time outside the matter; using equation (\ref{Volume}) and in particular 
$V = 4 \pi R |x_{\tilde{a}}| \, |x_b|^2$ we obtain the following spectrum of the volume operator
\begin{eqnarray}
\hat{V} | \mu_{\tilde{a}}, \mu_b \rangle= 4 \pi R \, |\hat{x}_{\tilde{a}}| \, |\hat{x}_b|^2 | \mu_{\tilde{a}}, \mu_b \rangle = 4 \pi R L_b^2  \, |\mu_{\tilde{a}}| \, |\mu_b|^2  |\mu_{\tilde{a}}, \mu_b \rangle.
\end{eqnarray}
Now we show that the operator $\frac{1}{\mbox{det(E)}} = \frac{1}{\sqrt{h}} \sim
 \frac{1}{|{\tilde{a}}| \, |b|^2}$ does not diverge in the quantum theory and so we don't  have any singularity.
 
Using extensively the relation (\ref{bsua}) we can define the classical and so the 
quantum operator $1/\mbox{det}(E)$ (we promote the Poisson Brackets to commutators) \cite{M2}. 
The spectrum of this operator is
\begin{eqnarray}
\widehat{\frac{1}{\mbox{det(E)}}} |\mu_{\tilde{a}}, \mu_b \rangle \, = \, \frac{2^6 \, 3^{15}}{L^2} \, 
                               |\mu_{\tilde{a}}|^5 \, |\mu_b|^6 \, [|\mu_b -1|^{\frac{1}{2}} -|\mu_b|^{\frac{1}{2}}]^{12} \, 
                               ||\mu_{\tilde{a}} -1|^{\frac{1}{3}} - |\mu_{\tilde{a}}|^{\frac{1}{3}}|^9 \, 
                               [|\mu_b -1|^{\frac{2}{3}} - |\mu_b|^{\frac{2}{3}}]^6,
 \end{eqnarray}
and we can see that it is upper bounded. 

Now we study the singularity problem from the matter point of view and so we report
the famous and important result of loop quantum cosmology because inside the matter, 
as said in the first section, space-time is exactly a Friedmann cosmological model with $k=1$, 
the operator that we study is the quantum analog of $1/|x_a|$ which we define
in quantum theory using the quantization of the classical Poisson bracket (\ref{Poisson.Volume}). 
In this way we obtain (for $\gamma_a = 1$) the operator
\begin{eqnarray}
 \widehat{\frac{1}{|x_a|}} = \frac{1}{2 \pi l_p^2 l_0^{\frac{2}{3}}} \left(
 \hat{U}^{-1}_{\gamma_a} \left[ \hat{V}_{in}^{\frac{1}{6}} , \hat{U}_{\gamma_a} \right] \right)^2. 
\end{eqnarray}
Using the volume operator $\widehat{V}_{in} = \mathcal{V}(\chi_0) \widehat{|x_a|}^3$ 
and the eigenvalue equation 
\begin{eqnarray}
\widehat{V}_{in} |\mu_a, \mu_{\phi} \rangle = \mathcal{V}(\chi_0) |\mu_a|^3 |\mu_a, \mu_{\phi}\rangle
\end{eqnarray}
 we obtain the action of the classical singular operator on the basis states 
  \begin{eqnarray}
 \widehat{\frac{1}{|x_a|}} \, | \mu_a, \mu_{\phi} \rangle = \sqrt{\frac{2}{ \pi l_p^2}}
 \left( | \mu_a |^{\frac{1}{2}} - |\mu_a -1|^{\frac{1}{2}}\right)^2 \, |
 \mu_a, \mu_{\phi} \rangle. 
\end{eqnarray}
We can see that the spectrum is bounded from below and so there is no singularity 
in the quantum theory also also in presence of matter.

\subsection{The Hamiltonian Constraint in Quantum Theory}

Now we want to resolve the Hamiltonian constraint in all the space time inside the horizon 
of the black hole in the case of dust matter. In fact only in this case we can match the
region inside the matter with the Kantowski-Sachs space time outside.  \\
The solutions of the Hamiltonian constraint are in the $\mathcal{C}^{\star}$ space that 
is the dual of the dense subspace $\mathcal{C}$ of the kinematical space $\mathcal{H}$.
A generic element of this space for our system which consists of two parts, a region where 
is the matter localized and an exterior region, is of the form 
\begin{eqnarray}
\langle \psi | = \sum_{\mu} \psi(\alpha, \beta) \langle \alpha, \beta|.
\end{eqnarray}
where the variables "$\alpha$" and "$\beta$" are the eigenvalues of the operators 
"$\widehat{x}_a$" and "$\widehat{x}_{\phi}$" inside the matter and "$x_b$" and "$x_{\tilde{a}}$" 
outside the matter. 
The constraint equation $\hat{H} |\psi \rangle = 0$ is now interpreted as an equation in the dual space $\langle \psi | \hat{H}^{\dag} = 0$;
from this equation we can derive a relation for the coefficients $\psi(\alpha, \beta)$. 
The Hamiltonian constraint that we must impose to define the physical space is 
\begin{equation}
\widehat{H} =  \left\{ \begin{array}{cc} {\widehat{H}}_{in} \hspace{0.5cm} \mbox{inside the matter} \\
                                                                   {\widehat{H}}_{out} \hspace{0.5cm} \mbox{outside the matter}                                                           \end{array}\right.
\end{equation}
Where $\widehat{H}_{in}$ and $\widehat{H}_{out}$ are define
\begin{eqnarray}
&& H_{in} = -\Bigg(\frac{p_a^2}{8} \, \frac{1}{|x_a|} + \frac{2}
{\mathcal{V}^{1/3}(\chi_0)} V^{1/3} \Bigg) + \frac{16 \pi G_N}{3} H_\phi(a),
      \nonumber \\
&& H_{out}  = \frac{G_N \, p_{{\tilde{a}}}^2}{2 R} \, \frac{|x_{\tilde{a}}|}{\,\, x_b^2}
            - \frac{G_N \, p_{\tilde{a}} \, p_b}{R} \, \frac{\mbox{sgn}(x_b) \, \, \mbox{sgn}(x_{\tilde{a}})}
            {|x_b|} 
            - \frac{R}{2 G_N} \, |x_{\tilde{a}}|
\label{Hamiltonian.1}
\end{eqnarray}
Now we quantize this Hamiltonian constraint. As we know, the operators $p_{\tilde{a}}$,
$p_b$, $p_a$ and $p_{\phi}$ don't exist in our quantum representation and so we choose the following alternative representation for the operators $p_{\tilde{a}}^2$, $p_{\tilde{a}} \, p_b$, 
$p_a^2$ and $p_{\phi}$. The first two operators were obtained in reference \cite{M2} starting
from the classical relations 
\begin{eqnarray}
&& p_a^2 = \frac{L_a^4}{(8 \pi G_N)^2} \mbox{lim}_{\gamma_a \rightarrow 0} \left( \frac{2 - U_{\gamma_a} - U_{\gamma_a}^{-1}}{\gamma_a^2} \right), \nonumber \\
&& p_a \, p_b = \frac{L_a^2 \, L_b}{2(8 \pi G_N)^2} \, \mbox{lim}_{\gamma_a, \gamma_b \rightarrow 0} 
            \Bigg[ \Bigg(\frac{U_{\gamma_a} + U_{\gamma_b} - U_{\gamma_a} \, U_{\gamma_b}  -1}{\gamma_a \, \gamma_b}\Bigg) + \Bigg(\frac{U_{\gamma_a}^{-1} + U_{\gamma_b}^{-1} - U_{\gamma_a}^{-1} \, U_{\gamma_b}^{-1}  -1}{\gamma_a \, \gamma_b}\Bigg)\Bigg], \nonumber \\
            &&
            \label{pi}
\end{eqnarray}
the other two operators can be define using the classical expressions  
 \begin{eqnarray}
&& p_a^2 = \frac{L_a^2}{\gamma_a} \mbox{lim}_{\gamma_a \rightarrow 0} \left( \frac{2 - U_{\gamma_a} - U_{\gamma_a}^{-1}}{\gamma_a^2} \right), \nonumber \\
&& p_{\phi} = \frac{L_{\phi}^4 \, L_b}{8 \pi G_N} \, \mbox{lim}_{\gamma_{\phi}, \rightarrow 0} 
            \Bigg(\frac{2 - U_{\gamma_{\phi}} - U_{\gamma_{\phi}}^{-1}}{\gamma_{\phi}^2} \Bigg). 
\end{eqnarray}
(we have a physical interpretation setting $\gamma_{\tilde{a}} = \gamma_{\phi} = \gamma_a = \gamma_b = l_F / L_{phys}$, where $L_{Phys}$ is the characteristic size of the system and $l_F$ is a fundamental length scale).

Now we are ready to construct the correct operators $\widehat{H}_{in}$ and $\widehat{H}_{out}$
in terms of the space volume and the momentum analog of the holonomy variable of
loop quantum gravity. 
From the paper \cite{M2} we report 
\begin{eqnarray}
\widehat{H}_{out} & = & \frac{1}{32 \pi^2 G_N R^2 \gamma_a^2 \gamma_b^4} \left[ 2 - \hat{U}_{\tilde{a}} - \hat{U}_{{\tilde{a}}}^{-1} \right]  \, 
   \left( \hat{U}_{b}^{-1} \left[ \hat{V}^{\frac{1}{4}} , \hat{U_b} \right] \right)^4 \nonumber \\
    & + & \frac{3^6}{2^{11} \pi^5 R^4 L^4 G_N \gamma_b^5 \gamma_a^7} \,   \Bigg[\Big( \frac{\hat{U}_{\tilde{a}} + \hat{U}_b - \hat{U}_{\tilde{a}} \, \hat{U}_b  -1}{2}\Big)+ \Big( \frac{\hat{U}_{\tilde{a}}^{-1} + \hat{U}_b^{-1} - \hat{U}_{\tilde{a}}^{-1} \, \hat{U}_b^{-1}  -1}{2}\Big)\Bigg]  \nonumber \\
    &&  
    \left( \hat{U_b}^{-1} \left[ \hat{V}^{\frac{1}{4}} , \hat{U_b} \right] \right)^4
    \, \left( \hat{U_{\tilde{a}}}^{-1} \left[ \hat{V}^{\frac{1}{3}} , \hat{U_{\tilde{a}}} \right] \right)^3 
    \, \left( \hat{U_b}^{-1} \left[ \hat{V}^{\frac{1}{3}} , \hat{U_b} \right] \right) ^3   + \nonumber \\
    & - & \frac{1}{8 \pi G_N L^2 \gamma_b^2} \left( \hat{U_b}^{-1} \left[ \hat{V}^{\frac{1}{2}} , \hat{U_b} \right]
     \right) ^2.
     \label{Hconstraint}
            \end{eqnarray}
 On the other side the operator ${\widehat{H}}_{in}$ is
 \begin{eqnarray}
 \widehat{H}_{in} = - \Bigg[ \frac{1}{2 \pi^2 \mathcal{V}(\chi_0) \gamma_a^4} \left(2 - \hat{U}_{a} - \hat{U}_{{a}}^{-1} \right)  \, \left( \hat{U}_{b}^{-1} \left[ \hat{V}^{\frac{1}{6}} , \hat{U_b} \right] \right)^2  + \frac{2}{\mathcal{V}(\chi_0)^{1/3}} \, \hat{V}^{\frac{1}{3}} \Bigg] + \frac{16 \pi G_N}{3} \, \widehat{H}_{\phi}(x_a) 
         \label{Hconstraintin}
            \end{eqnarray}
            
Now we resolve the Hamiltonian constraint. As we said at the beginning of this section the solutions of the Hamiltonian constraint is obtained in the $\mathcal{C}^{\star}$.
A generic element of this space for the space-time inside the matter is 
\begin{eqnarray}
\langle \psi | = \sum_{\mu_{\tilde{a}}, \mu_b} \psi(\mu_{\tilde{a}}, \mu_b) \langle \mu_{\tilde{a}}, \mu_b|.
\end{eqnarray}
In \cite{M2} we obtained an equation for the coefficients $\psi(\mu_{\tilde{a}}, \mu_{b})$ 
\begin{eqnarray}
&& \hspace{-0.7cm}[2 \alpha(\mu_{\tilde{a}}, \mu_b) - 2 \beta(\mu_{\tilde{a}}, \mu_b) + \gamma(\mu_{\tilde{a}}, \mu_b)] \, \psi(\mu_{\tilde{a}}, \mu_b) -
[\alpha(\mu_{\tilde{a}} + \gamma_{\tilde{a}}, \mu_b) - \beta(\mu_{\tilde{a}} + \gamma_{\tilde{a}}, \mu_b)] \, \psi(\mu_{\tilde{a}} + \gamma_{\tilde{a}}, \mu_b) + \nonumber \\
&& - [\alpha(\mu_{\tilde{a}} - \gamma_{\tilde{a}}, \mu_b) + \beta(\mu_{\tilde{a}} - \gamma_{\tilde{a}},        \mu_b)] \, \psi(\mu_{\tilde{a}} - \gamma_{\tilde{a}}, \mu_b) +
       \beta(\mu_{\tilde{a}}, \mu_b + \gamma_b) \,\psi(\mu_{\tilde{a}}, \mu_b + \gamma_b) + \nonumber \\
       && - \beta(\mu_{\tilde{a}}, \mu_b - \gamma_b) \, 
       \psi(\mu_{\tilde{a}}, \mu_b -\gamma_b) +
       \beta(\mu_{\tilde{a}}+\gamma_{\tilde{a}}, \mu_b + \gamma_b) \, \psi(\mu_{\tilde{a}}
       +\gamma_{\tilde{a}}, \mu_b+\gamma_b) \nonumber \\
       && - \beta(\mu_{\tilde{a}} - \gamma_{\tilde{a}}, \mu_b -\gamma_b) \, \psi(\mu_{\tilde{a}} 
       - \gamma_{\tilde{a}}, \mu_b -\gamma_b)  = 0 
      \label{difference}
\end{eqnarray}
where the function $\alpha, \beta, \gamma$, always following \cite{M2}, are 
\begin{eqnarray}
 \alpha(\mu_{\tilde{a}}, \mu_b) & = &\frac{L^2}{8 \pi^2 R G_N \gamma_b^4 \gamma_a^2} \,
      \Big(|\mu_{\tilde{a}}|^{\frac{1}{4}} |\mu_b - \gamma_b|^{\frac{1}{2}} -|\mu_{\tilde{a}}|^{\frac{1}{4}} |\mu_b|^{\frac{1}{2}} \Big)^2,  \nonumber \\
 \beta(\mu_{\tilde{a}}, \mu_b) & = & - \frac{L^2}{2(8 \pi)^2 G_N R \gamma_b^5 \gamma_a^7} \, 
       \Big(|\mu_{\tilde{a}}|^{\frac{1}{4}} |\mu_b - \gamma_b|^{\frac{1}{2}} -|\mu_{\tilde{a}}|^{\frac{1}{4}} |\mu_b|^{\frac{1}{2}} \Big)^4 \nonumber \\
    && \hspace{2.3cm}   \Big(|\mu_{\tilde{a}} - \gamma_a|^{\frac{1}{3}} |\mu_b|^{\frac{2}{3}} -|\mu_b|^{\frac{1}{3}} |\mu_b|^{\frac{2}{3}} \Big)^3  \nonumber\\
     &&\hspace{2.3cm} \Big(|\mu_{\tilde{a}}|^{\frac{1}{3}} |\mu_b - \gamma_b|^{\frac{2}{3}} -|\mu_{\tilde{a}}|^{\frac{1}{3}} |\mu_b|^{\frac{2}{3}} \Big)^3, \nonumber \\        
     \gamma(\mu_{\tilde{a}}, \mu_b) & = & \frac{R}{2 G_N \gamma_b^2} \, 
     \Big(|\mu_{\tilde{a}}|^{\frac{1}{2}} |\mu_b - \gamma_b| - |\mu_{\tilde{a}}|^{\frac{1}{2}} |\mu_b|   \Big)^2.     
\end{eqnarray}
Now we evaluate the action of $\hat{H}_{in}$ on the states $|\mu_a, \mu_{\phi} \rangle$  
\begin{eqnarray}
\hat{H}_{in} \, |\mu_a, \mu_{\phi} \rangle & = & -\frac{L}{2 \gamma_a^4} \Big(|\mu_a -\gamma_a|^{\frac{1}{2}} - |\mu_a|^{\frac{1}{2}} \Big)^2 \, 
\Big( 2 \, |\mu_a, \mu_{\phi} \rangle - |\mu_a - \gamma_a, \mu_{\phi}Ê\rangle
 - |\mu_a + \gamma_a, \mu_{\phi} \rangle \Big) \nonumber \\
 &&- 2 L |\mu_a| \, |\mu_a, \mu_{\phi} \rangle + \frac{16 \pi G_N}{3} \, \widehat{H}_{\phi}(x_a) 
 |\mu_a, \mu_{\phi} \rangle
  \end{eqnarray}
The solution of the Hamiltonian constraint is in the dual space of the dense subspace of the
kinematical space $\mathcal{H}$ and so a generic element of this space, as we said at the
beginning of this section, is $\langle \psi | = \sum_{\mu_a, \mu_{\phi}} \psi(\mu_a, \mu_{\phi}) \langle \mu_a, \mu_{\phi}|$. The equation for the coefficients $\psi(\mu_a, \mu_{\phi})$ is
\begin{eqnarray}
&&\hspace{-0.7cm} \alpha(\mu_a) \psi(\mu_a, \mu_{\phi}) +  \beta(\mu_a + \gamma_a) \psi(\mu_a+ \gamma_a, \mu_{\phi}) + \beta(\mu_a - \gamma_a) \psi(\mu_a - \gamma_a, \mu_{\phi}) = 
 - \frac{16 \pi G_N}{3} \hat{H}_{\phi}(a) \psi(\mu_a, \mu_\phi), \nonumber \\
 &&\label{difference2}
\label{intern.wave}
\end{eqnarray}
where the functions of the eigenvalues $\alpha(\mu_a)$ and $\beta(\mu_a)$ are 
\begin{eqnarray}
\alpha(\mu_a) & = & - \frac{L}{\gamma_a^4} \Big(|\mu_a - \gamma_a|^{\frac{1}{2}} - |\mu_a|^{\frac{1}{2}} \Big)^2  - 2 L |\mu_a| \nonumber \\
\beta(\mu_a) & = & \frac{L}{2 \gamma_a^4} \Big(|\mu_a - \gamma_a|^{\frac{1}{2}} - |\mu_a|^{\frac{1}{2}} \Big)^2  
 \end{eqnarray}
and $\hat{H}_{\phi}(a) \psi(\mu_a, \mu_\phi) = \sum_{\mu_a^{'}, \mu_{\phi}^{'}} \psi(\mu_a^{'}, \mu_{\phi}^{'}) \langle \mu_a^{'}, \mu_{\phi}^{'} | \hat{H}_{\phi}(a) | \mu_a, \mu_{\phi} \rangle$.\\
At this point we have the wave solution inside and outside the matter for the gravitational
collapse and we can give an interpretation for the equation (\ref{difference})
and (\ref{difference2}). Both equations are difference equations and the physical states 
are combinations of a countable number of  components of the form
$\psi(\mu + n \gamma, \nu + m \delta) |\mu + n \gamma, \nu + \delta \rangle$ ($\gamma, \delta \sim l_P/L_{Phys} \sim 1$; $\mu = \mu_a$ or $\mu = \mu_{\tilde{a}}$ and $\gamma = \gamma_a$ or 
$\gamma = \gamma_{\tilde{a}}$, $\nu = \mu_{\phi}$ or $\nu = \mu_b$ inside or outside the matter). 
Outside the matter any component corresponds to a particular value of the volume; so we can interpret $\psi(\mu_{\tilde{a}} + \gamma_{\tilde{a}}, \mu_b + \gamma_b)$ as the function of the Black Hole inside the horizon at the time $\mu_b + \gamma_b$, if we interprete $b$ as the time and $\tilde{a}$ as the space partial observable that defines the quantum fluctuations around the Schwarzschild solution.  In the same way, inside the matter, we can interpret the function 
$\psi(\mu_a + \gamma_a, \mu_{\phi} + \gamma_{\phi})$ as the wave function of the matter 
at the time $\mu_a + \gamma_a$. In the next section we impose the boundary condition on
the area operator from inside and from outside the matter view and we obtain a single time
coordinate. 

\subsection{Boundary Condition and Time Arrow}
At this point we can impose the boundary condition inspired from the classical 
condition on the inside and outside area of the star. In particular we impose that
the operator area spectrum which define the surface of the star from the point 
of view of the inside region is identically to the area operator  spectrum from the
point of view of the region outside the matter.
The operator area are 
\begin{eqnarray}
&& \widehat{A}_{in} = 4 \pi \widehat{|x_a|}^2 \mbox{sin}(\chi_0), \nonumber \\
&& \widehat{A}_{out} = 4 \pi \widehat{|x_b|}^2.
\end{eqnarray}
The spectrum of the two operators can be obtained from the bases (\ref{base.in}) and (\ref{base.out})
\begin{eqnarray}
&& \widehat{A}_{in} |\mu_a, \mu_{\phi} \rangle = 4 \pi \widehat{|x_a|}^2 \mbox{sin}^2(\chi_0) 
     \, |\mu_a, \mu_{\phi} \rangle = 4 \pi |\mu_a|^2 \mbox{sin}^2(\chi_0) \, |\mu_a, \mu_{\phi} \rangle, 
      \nonumber \\
&& \widehat{A}_{out} |\mu_{\tilde{a}}, \mu_{b} \rangle= 4 \pi \widehat{|x_b|}^2 \, |\mu_{\tilde{a}}, \mu_{b} \rangle = 4 \pi |\mu_b|^2 \, |\mu_{\tilde{a}}, \mu_{b} \rangle. 
\end{eqnarray}
At this point we identify the inside and outside spectrum and we obtain a relation between
the inside and outside eigenvalues $\mu_a$ and $\mu_b$
\begin{eqnarray}
&& |\mu_a|^2 \mbox{sin}^2(\chi_0) =  |\mu_b|^2.
\end{eqnarray}
From this relation we can see that $\mu_a \sim \mu_b$ and so if we define "$a$" and "$b$" 
as time coordinates for the inside and outside space-time the boundary condition imply 
that we have only one time coordinate in all space-time.

\subsection{Quantum Dust Matter}
In this subsection we quantize the simply Hamiltonian for the dust matter.
In particular we follow the quantization program of reference \cite{AFW} and apply
the  non unitary equivalent quantization to the isotropic and  homogeneous dust matter.

We give the form of the Hamiltonian constraint in comoving coordinates. 
In this case the energy
tensor is homogeneous and so it is independent from the spatial section coordinates. 
In comoving coordinates the supermomentum is automatically zero and the classical 
hamiltonian becomes 
\begin{eqnarray}
H_{\phi} = |p_{\phi}|.
\end{eqnarray}
As $p $ is positive as shown in \cite{Brown}, we can take $H = p_{\phi}$.
Now using the classical variables (\ref{UgammaphiDust}), we can obtain the quantum 
operator $H_{\phi}$
\begin{eqnarray}
\widehat{H}_{\phi} = \widehat{p}_{\phi} =\frac{L_{\phi}}{G_N} \Bigg(\frac{\widehat{U}_{\phi} - \widehat{U}_{\phi}^{-1}}{2 i \gamma_{\phi}}\Bigg).
\end{eqnarray}
The action of this operator on the state $|\mu_a, \mu_{\phi} \rangle$ is
\begin{eqnarray}
\widehat{H}_{\phi} |\mu_a, \mu_{\phi} \rangle = \frac{L_{\phi}}{2 i \gamma_{\phi}G_N} 
\Big(|\mu_a, \mu_{\phi} - \gamma_{\phi} \rangle - |\mu_a, \mu_{\phi} + \gamma_{\phi}\rangle\Big)
\end{eqnarray}

\subsection{Quantum Scalar Field}

In this section we quantize the scalar field in the same way of the dust matter sector and 
of the gravity sector;
in particular we follow the quantization program of reference \cite{AFW} and apply
the  non unitary equivalent quantization to an isotropic and  homogeneous scalar field.\\
We remember that the scalar field Hamiltonian constraint operator from (\ref{Scal.Con}) is 
\begin{eqnarray}
\widehat{H}_{\phi} = \frac{\hat{p}_{\phi}^2}{2} \, \widehat{\frac{1}{|x_a|^3}}
      + \widehat{U(\phi)} \, \widehat{|x_a|}^3 .
      \label{Scal.Con.1}
\end{eqnarray}
Now using the relation (\ref{Ugammaphi}), the Poisson brackets (\ref{Poisson.Volume1}) and the scalar field analog of the momentum operator (\ref{pi}), we can write the operator (\ref{Scal.Con.1}) in terms of $V_{in}$, $U_{\gamma_a}$ and
the scalar field couple $(\phi, U_{\gamma_{\phi}})$,
\begin{eqnarray}
\widehat{H}_{\phi} =  \Bigg[ \frac{1}{2 G_N l_p^2 \pi^2 \mathcal{V}(\chi_0) \gamma_a^6 \gamma_{\phi}^2}  \left(2 - \hat{U}_{\gamma_{\phi}} - \hat{U}_{\gamma_{\phi}}^{-1} \right)  \, \left( \hat{U}_{\gamma_a}^{-1} \left[ \hat{V}^{\frac{1}{6}}_{in} , \hat{U}_{\gamma_a} \right] \right)^6  + 
 \frac{1}{\mathcal{V}(\chi_0)} \, \widehat{U}(\phi) \, \hat{V}_{in}\Bigg].
\label{Scal.Con.2}
\end{eqnarray}
At this point we can apply the operator (\ref{Scal.Con.1}) on the basis $|\mu_a, \mu_{\phi} \rangle$ 
and obtain 
\begin{eqnarray}
&& \hspace{-0.8cm}\widehat{H}_{\phi} |\mu_a, \mu_{\phi} \rangle  =  \frac{L_a^3}{2 G_N l_p^2 \pi^2 \mathcal{V}(\chi_0) \gamma_a^6 \gamma_{\phi}^2} \, \big(|\mu_a - \gamma_a|^{\frac{1}{2}} - |\mu_a|^{\frac{1}{2}}\big)^6 
\Big(2 |\mu_a, \mu_{\phi} \rangle - |\mu_a, \mu_{\phi} - \gamma_{\phi}  \rangle - |\mu_a, \mu_{\phi} + \gamma_{\phi} \rangle \Big) \nonumber \\
&& \hspace{1.3cm}+ L_a^3 \, |\mu_a|^3 \, U(\mu_{\phi}) \, |\mu_a, \mu_{\phi} \rangle.
\label{scalar}
\end{eqnarray}
Using this relation we obtain the complete quantum solution of the Hamiltonian constraint
inside the matter 
\begin{eqnarray}
&& \alpha(\mu_a, \mu_{\phi}) \psi(\mu_a, \mu_{\phi}) +  \beta(\mu_a + \gamma_a) \psi(\mu_a + \gamma_a, \mu_{\phi}) + \beta(\mu_a - \gamma_a) \psi(\mu_a - \gamma_a, \mu_{\phi}) \nonumber \\
&& + \gamma(\mu_a) \psi(\mu_a, \mu_{\phi} + \gamma_{\phi}) + 
 \gamma(\mu_a) \psi(\mu_a, \mu_{\phi} - \gamma_{\phi}) = 0 
\label{difference2.2}
\end{eqnarray}
where the functions $\alpha(\mu_a, \mu_{\phi})$, $\beta(\mu_a)$ and $\gamma(\mu_a)$ are 
\begin{eqnarray}
\alpha(\mu_a, \mu_{\phi}) & = & - \frac{L}{\gamma_a^4} \Big(|\mu_a - \gamma_a|^{\frac{1}{2}} - |\mu_a|^{\frac{1}{2}} \Big)^2  - 2 L |\mu_a|  \nonumber \\
&& + \frac{16 l^3}{3 l_p^2 \pi \gamma_a^6 \gamma_{\phi}^2} \Big(|\mu_a - \gamma_a|^{\frac{1}{2}} - |\mu_a|^{\frac{1}{2}} \Big)^6 + \frac{16 \pi G_N L^3}{3} |\mu_a|^3 U(\mu_{\phi})
\nonumber \\
\beta(\mu_a) & = & \frac{L}{2 \gamma_a^4} \Big(|\mu_a - \gamma_a|^{\frac{1}{2}} - |\mu_a|^{\frac{1}{2}} \Big)^2  \nonumber \\
\gamma(\mu_a) & = & - \frac{8 L^3}{3 l_p^2 \pi \gamma_a^6 \gamma_{\phi}^2} 
\Big(|\mu_a - \gamma_a|^{\frac{1}{2}} - |\mu_a|^{\frac{1}{2}} \Big)^6
 \end{eqnarray}
In this case we cannot mach the solution inside the matter with the Schwarzschild solution
outside the matter. In fact for the scalar matter the pressure it is not  equal to zero
as for the Schwarzschild solution or the Kantowski-Sachs symmetric solution. We can 
however match the scalar matter solution with the Vaydia  solution.

\section*{Conclusions}

In this work we have applied the non Schr$\ddot{\mbox{o}}$dinger
quantization procedure used in the previous papers \cite{work1}, 
\cite{M2} onthe black hole singularity and in the work 
of V. Husain and O. Winkler on quantum cosmology. This quantum 
formalism was introduced by Halvorson \cite{Fonte.Math} and also
by A. Ashtekar, S. Fairhust and J. Willis \cite{AFW} in a very clear 
mathematical form. 
In this paper we have studied the gravitational collapse inside the 
horizon or, in other words, when all the matter has passed the 
horizon of the black hole. In this particular configuration we have subdivided
the space-time inside the horizon in two regions, one where the dust matter 
is localized and the other where the space time is the Kantowski-Sachs 
space-time. We have studied also the quantum theory when the matter
is an homogeneous, isotropic scalar field but in this case the pressure 
is not zero and does not give a matching with the space time outside the 
matter. 

The principal results of our model are the following.

The first one 
is the absence of singularity in the space-time region where the matter
is localized. In fact in the particular case of dust matter but also in the
case of scalar matter the model is analogous to a cosmological model
where the singularity is absent in the quantum theory \cite{Boj}.   
On the other side the space-time outside the matter, which is 
a Kantowski-Sachs space time \cite{KS} with space topology  
$\mathbf{R} \times \mathbf{S}^2$ and which contains the 
Schwarzschild metric inside the  horizon as a particular classical
solution, is singularity free. The other interesting point is that
the space-time can be extended beyond the singularity. This
is correct for the region which contains the matter and
the vacuum region.

The other interesting point is the quantum match between 
the two regions. When we impose the classical matching between the
area of the 2-sphere $\mathbf{S}_2$ inside and outside the matter 
at the quantum level we obtain the interesting result that only a 
temporal coordinate survive in all the space time. In fact we start 
with two coordinates inside the matter "$a$" and "$\phi$" that we
interpret as time ($a$) and dust matter ($\phi$) and two coordinates
outside the matter, but inside the horizon, that are 
the time coordinate "$b$" and the gravitational coordinate "$\tilde{a}$". 
When we match the area spectrum  of the 2-sphere inside the
matter $b^2 (t)  (\sin^2 \theta d\phi^2 + d \theta^2)$ and the area 
spectrum of the 2-sphere outside the matter 
$ a^2 (t) \sin^2 \chi_0 (\sin^2 \theta d\phi^2 + d \theta^2)$ we obtain 
only one time coordinate. This relation is the quantum 
version of the classical matching $a \sin \chi_0 = b$: for
any eigenvalue $\mu_a$ of the operator $"\hat{a}"$ and for
any eigenvalue $\mu_b$ of $"\hat{b}"$ we have $|\mu_a| \sin \chi_0 = |\mu_b|$.



\section*{Acknowledgements}
We are grateful to Carlo Rovelli and Eugenio Bianchi for many important and clarifying discussion about this work. This work is supported in part by a grant from the Fondazione Angelo Della Riccia.

\end{document}